\documentclass[aps,prb,showpacs,twocolumn,floats]{revtex4}
\usepackage{amssymb}

\usepackage{amsmath}
\usepackage{graphicx,psfig}
\usepackage{dcolumn}
\usepackage{bm}

\input{tcilatex}

\begin{document}

\title{Theory of magnetic deflagration}
\author{D. A. Garanin and E. M. Chudnovsky}
\affiliation{Department of Physics and Astronomy, Lehman College,
City University of New York \\ \mbox{250 Bedford Park Boulevard
West, Bronx, New York 10468-1589, U.S.A.}}
\date{25 April 2007}

\begin{abstract}
Theory of magnetic deflagration (avalanches) in crystals of molecular
magnets has been developed. The phenomenon resembles the burning of a
chemical substance, with the Zeeman energy playing the role of the chemical
energy. Non-destructive reversible character of magnetic deflagration, as
well as the possibility to continuously tune the flammability of the crystal
by changing the magnetic field, makes molecular magnets an attractive toy
system for a detailed study of the burning process. Besides simplicity, new
features, as compared to the chemical burning, include possibility of
quantum decay of metastable spin states and strong temperature dependence of
the heat capacity and thermal conductivity. We obtain analytical and
numerical solutions for criteria of the ignition of magnetic deflagration,
and compute the ignition rate and the speed of the developed deflagration
front.
\end{abstract}
\pacs{75.50.Xx, 76.60.Es, 82.33.Vx}
\maketitle


\section{Introduction}

Recently, it has been observed that molecular magnets exhibit explosive
relaxation towards thermal equilibrium that resembles propagation of a flame
through a flammable chemical substance. \cite{suzetal05prl} Theory of this
effect is the subject of this paper.

Crystals of molecular magnets first attracted attention of physicists after
it was demonstrated \cite{sesgatcannov93nat} that individual molecules
inside such crystals behave as superparamagnetic particles. \cite
{chutej06book} Due to large molecular spin (e.g., $S=10$ for Mn-12 and Fe-8
molecular magnets) and high magnetic anisotropy, spin-up and spin-down
states of many molecular magnets are separated by a large energy barrier.
Consequently, unlike conventional paramagnets, molecular magnets are
characterized by a macroscopic time of thermal relaxation between spin-up
and spin-down states. Similarly large times are needed for quantum
transitions between these states to occur, which allows one to speak about
quantum tunneling of the magnetic moment. \cite{chutej98book} Due to this
effect, molecular magnets exhibit spectacular staircase magnetization curve.
\cite{frisartejzio96prl} It has been known for some time that the
low-temperature magnetic relaxation in crystals of molecular magnets can
occur via two mechanisms. The first, slow mechanism involves random thermal
and quantum spin transitions at unrelated spatial points. Such transitions
influence each other only through weak long-range dipolar fields associated
with the magnetic moments of the molecules. \cite{feralo05prb} The resulting
relaxation lasts macroscopic times. This allows one to study transition
rates by simply measuring the time dependence of the macroscopic
magnetization of the crystal.

The second mechanism of relaxation -- magnetic avalanches -- corresponds to
the abrupt reversal of the magnetization when a sufficiently large crystal
is placed in a large magnetic field opposite to its magnetic moment. \cite
{fometal97prl,paupark95kluwer,baretal99prb} The avalanche was long believed
to be a thermal runaway in which the Zeeman energy released by the relaxing
molecules gets transformed into heat that generates transitions in the
neighboring molecules and accelerates the total energy release. Such a
relaxation that typically occurs in a millisecond time was long considered a
nuisance as it often interfered with experimental studies of spin tunneling.
More recently, it was realized through time-resolved local measurements \cite
{suzetal05prl} that magnetic avalanches resemble propagation of a flame --
deflagration -- in which the role of the chemical energy stored in a
flammable substance is played by the Zeeman energy. Due to quantum tunneling
between spin states that occurs at discrete values of the magnetic field,
magnetic deflagration also exhibits quantum features. \cite{heretal05prl}

Experiments performed to date have established with certainty that magnetic
avalanches in crystals of molecular magnets correspond to the propagation of
a narrow front of the magnetization reversal. The analogy with burning of a
flammable chemical substance has been confirmed by the study of the
dependence of the flame speed on the energy barrier. In crystals of
molecular magnets, the latter can be continuously tuned by the magnetic
field. For the study of deflagration this tunability of the barrier, as well
as the reversible nature of the magnetic burning, provides a great advantage
over irreversible burning of a chemical substance with a fixed energy
barrier. Thus a detailed study of magnetic deflagration can answer important
questions of the theory of combustion and detonation. \cite{gla96book} There
are also novel features that are absent in conventional combustion. They
include a very strong temperature dependence of the specific heat and
thermal conductivity of molecular magnets at low temperature \cite
{gometal98prb,gomnovnunrap01jmmm,fometal99prb} and the possibility of
magnetization reversal via quantum tunneling.

In this paper we intend to answer the following questions:

\begin{itemize}
\item  The critical combination of parameters (magnetic field, initial
temperature, and the size of the sample) that sets off the deflagration
process.

\item  The mode of instability.

\item  The time that elapses between bringing the system above the
deflagration threshold and the ignition of the deflagration process (the
ignition time).

\item  The temperature of the flame and the velocity of the deflagration
front.
\end{itemize}

We will show that the ignition of magnetic deflagration in molecular magnets
is very different from the ignition of magnetization reversal in
ferromagnets. The latter is dominated by the exchange interaction and begins
with the nucleation of a small critical nucleus of opposite magnetization
that spreads and occupies the entire sample. On the contrary, the magnetic
deflagration in a paramagnetic crystal of magnetic molecules begins as a
large-scale instability of a smooth temperature profile inside the sample
against formation of a rapidly moving deflagration front.

The structure of the paper is as follows. Properties of molecular magnets
will be discussed in Sec.\ \ref{Sec-MM}. The mechanism of thermal runaway in
a crystal of magnetic molecules will be analyzed in Sec.\ \ref{Sec-Runaway}.
Stability of the quasi-stationary temperature profile in a crystal of
molecular magnets will be analyzed in Sec.\ \ref{Sec-threshold}. The ignition
rate will be studied in Sec.\ \ref{Sec-Ignition-rate}. Structure and the
velocity of a developed deflagration front will be investigated in Sec.\ \ref
{Sec-Front}. Numerical illustrations of the deflagration process will be
given in Sec.\ \ref{Sec-Numerical}. Relevance of our results to experiment
and possible future directions of theory and experiment will be discussed in
Sec.\ \ref{Sec-Discussion}.

\section{Molecular magnets}

\label{Sec-MM}

\subsection{Magnetic bistability and spin tunneling}

\label{Sec-MM-Bistability}

\begin{figure}[t]
\unitlength1cm
\begin{picture}(11,5.5)
\centerline{\psfig{file=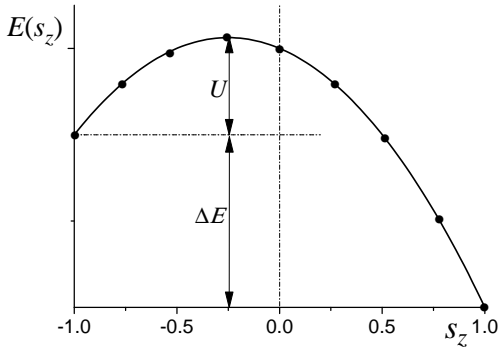,angle=-90,width=8cm}}
\end{picture}
\caption{Energy of a molecular magnet as function of $s_{z}$. Quantum energy
levels are shown by black circles for $S=4.$}
\label{Fig-Barrier}
\end{figure}

A single molecule of a molecular magnet can be described by the Hamiltonian
\begin{equation}
\mathcal{H}=-DS_{z}^{2}-g\mu _{B}H_{z}S_{z}+\mathcal{H}^{\prime },
\label{Ham}
\end{equation}
where $S$ is spin$,$ $D$ is the constant of the uniaxial $z$ anisotropy that
creates magnetic bistability, $H_{z}$ is the bias magnetic field, and $%
\mathcal{H}^{\prime }$ is a small part of the Hamiltonian that does not
commute with $S_{z}$ and is responsible for spin tunneling. If $S$ is large
(as, e.g., in Mn$_{12}$ and Fe$_{8})$, magnetic bistability can be
reasonably well described within the classical model with the energy that
depends on the classical vector $\mathbf{s=S}/S$ and has the form
\begin{equation}
E=-\left( s_{z}^{2}+2hs_{z}\right) U_{0}.  \label{EClassDef}
\end{equation}
Here
\begin{equation}
U_{0}=DS^{2},\qquad h\equiv \frac{g\mu _{B}H_{z}}{2DS}  \label{U0Def}
\end{equation}
are the zero-bias energy barrier and the reduced bias field. The dependence $%
E(s_{z})$ is shown in Fig.\ \ref{Fig-Barrier}. The spin-projection value
corresponding to the barrier between the two wells follows from $dE/ds_{z}=0$
and is given by $s_{z}^{(b)}=-h.$ The minima of $E$ and its value at the top
of the barrier are
\begin{equation}
E_{\pm }=-\left( 1\pm 2h\right) U_{0},\qquad E_{b}=h^{2}U_{0}.
\label{EpmbDef}
\end{equation}
Thus the values of the energy barriers for the molecules on the left and on
the right are given by $U_{\pm }=E_{b}-E_{\pm }=\left( 1\pm h\right)
^{2}U_{0}.$ In the case of $h>0,$ that we will consider throughout the
paper, $s_{z}=-1$ is a metastable minimum, whereas $s_{z}=1$ is the absolute
minimum of the energy. Below we will use $U_{-}\equiv U$,
\begin{equation}
U=\left( 1-h\right) ^{2}U_{0}.  \label{UDef}
\end{equation}
The energy difference between the two minima is given by
\begin{equation}
\Delta E=E_{-}-E_{+}=4hU_{0}.  \label{DeltaEDef}
\end{equation}

The noncommuting term $\mathcal{H}^{\prime }$ in Eq.\ (\ref{Ham}) gives rise
to resonance spin tunneling between the states at the two sides of the
barrier if the bias field satisfies the condition
\begin{equation}
g\mu _{B}H_{z}=kD,\qquad k=0,\pm 1,\pm 2,\ldots  \label{ResCond}
\end{equation}
This modifies the process of thermal activation of spins at low
temperatures. Off resonance, the spins have to be thermally activated all
the way up from the bottom of the metastable well to the top of the barrier.
On resonance, however, it is sufficient to be thermally activated up to the
energy level below the barrier where spin tunneling is sufficiently strong
to take the molecule to the other side of the barrier. This leads to the
resonance increase of the relaxation rate, see Fig.\ 7 of Ref. %
\onlinecite{garchu97prb}. On the phenomenological level, this effect can be
encapsulated into the effective barrier with dips at the resonance bias
fields given by Eq.\ (\ref{ResCond}). Since the exact form of the effective
barrier depends on the form of $\mathcal{H}^{\prime }$ that we do not
analyze in this paper, we will use for numerical work the fitting function
taken from experiments\cite{heretal96epl,baretal99prb} on Mn$_{12}$ and
replace Eq.\ (\ref{UDef}) by $U(h)=u(h)U_{0}$ with
\begin{equation}
u(h)=\left( 1-h\right) ^{2}-0.0806\left[ 1-\left| \sin \left( \pi \frac{g\mu
_{B}H_{z}}{D}\right) \right| \right] ^{2}.  \label{UQDef}
\end{equation}

\subsection{Magnetic relaxation and heat transfer}

At low temperatures
\begin{equation}
\frac{U}{k_{B}T}\equiv W\gg 1  \label{WDef}
\end{equation}
thermally activated transition of magnetic molecules over the barrier can be
described by the rate equations for the numbers of molecules in the left and
right wells, $n_{\pm },$ that satisfy $n_{+}+n_{-}=1.$ The equation for the
number of particles in the metastable well $n_{-}$ has the form
\begin{equation}
\dot{n}_{-}=\Gamma _{-+}n_{+}-\Gamma _{+-}n_{-}=-\Gamma \left[ n_{-}-n_{-}^{(%
\mathrm{eq})}\right] ,  \label{nminusEq}
\end{equation}
where $\Gamma =\Gamma _{+-}+\Gamma _{-+}.$ In accordance with the
detailed-balance condition
\begin{equation}
\frac{\Gamma _{+-}}{\Gamma _{-+}}=\frac{n_{+}^{(\mathrm{eq})}}{n_{-}^{(%
\mathrm{eq})}}=\exp \left( \frac{\Delta E}{k_{B}T}\right) ,\quad n_{-}^{(%
\mathrm{eq})}=\frac{1}{\exp \left( \frac{\Delta E}{k_{B}T}\right) +1}.
\label{DetailedBalance}
\end{equation}
Using $\Gamma _{+-}=\Gamma _{0}e^{-W}$ for $W\gg 1,$ one obtains
\begin{equation}
\Gamma =\Gamma _{0}e^{-W}\left[ 1+\exp \left( -\frac{\Delta E}{k_{B}T}%
\right) \right] ,  \label{GammaRes}
\end{equation}
where the second term square brackets describes transitions from the stable
well to the metastable well. In the strong-bias case, $\Delta E\gg k_{B}T,$
this term can be omitted. This yields simply $\Gamma =\Gamma _{0}e^{-W},$
and sets $n_{-}^{(\mathrm{eq})}=0$ (full burning).

When a magnetic molecule makes a transition from the metastable state $%
s_{z}=-1$ to the absolute energy minimum $s_{z}=1,$ the energy $\Delta E$ is
released. Thermalization of this energy leads to the temperature change $%
\Delta T=\Delta E/C_{\mathrm{ph}},$ where $C_{\mathrm{ph}}$ is the phonon
heat capacity per magnetic molecule. Other contributions to the specific
heat at low temperatures are considered small. The magnetic relaxation
creates a source in the heat conduction equation. Another term in this
equation is the divergence of the heat flow
\begin{equation}
\mathbf{q}=-k\nabla T,
\end{equation}
where $k$ is thermal conductivity. The full system of equations for the
temperature $T$ and the population of the metastable minimum $n_{-}$ has the
form
\begin{eqnarray}
\frac{\partial T}{\partial t} &=&\frac{1}{C_{\mathrm{ph}}}\nabla \cdot
k\nabla T-\frac{\Delta E}{C_{\mathrm{ph}}}\frac{\partial n_{-}}{\partial t}
\notag \\
\frac{\partial n_{-}}{\partial t} &=&-\Gamma \left[ n_{-}-n_{-}^{(\mathrm{eq}%
)}\right] .  \label{TnEqsDimensional}
\end{eqnarray}

An important feature of magnetic deflagration is strong temperature
dependence of the heat capacity and thermal conductivity at low
temperatures. As the temperatures before and behind the deflagration front
can differ by an order of magnitude, this effect cannot be neglected. The
phonon heat capacity $C_{\mathrm{ph}}$ has the form
\begin{equation}
C_{\mathrm{ph}}=Ak_{B}\left( \frac{T}{\Theta _{D}}\right) ^{\alpha },
\label{Cph3d}
\end{equation}
where $\alpha =3$ in three dimensions, $A$ is a numerical factor and $\Theta
_{D}$ is the Debye temperature. At low temperatures only acoustic phonons
are excited, whereas high-energy optical phonons are frozen out. Thus one
can use the $A$ value for the simple model of a crystal,\cite{kit63} $%
A=12\pi ^{4}/5\simeq 234$ that is in a good agreement with measurements\cite
{gometal98prb} on Mn$_{12}$. The thermal diffusivity
\begin{equation}
\kappa =k/C_{\mathrm{ph}}  \label{kappaDef}
\end{equation}
depends on the average mean free path of thermal phonons. At low
temperatures the main scattering mechanism is scattering on impurities, so
that (see Ref. \onlinecite{garlut92ap} and references therein)
\begin{equation}
\kappa \varpropto T^{-\beta },\qquad \beta =13/3.  \label{kappaPowerLaw}
\end{equation}
Accordingly the thermal conductivity behaves as
\begin{equation}
k\varpropto T^{-\gamma },\qquad \gamma =\beta -\alpha =4/3  \label{kPowerLaw}
\end{equation}
for $\alpha =3$.

The heat-conduction equation can be brought into a more elegant form by
choosing the phonon energy $\mathcal{E}$ as the dynamical variable. Using $%
C_{\mathrm{ph}}=d\mathcal{E}/dT$ one obtains
\begin{equation}
\frac{\partial \mathcal{E}}{\partial t}=\nabla \cdot \kappa \nabla \mathcal{E%
}-\Delta E\frac{\partial n_{-}}{\partial t}.  \label{UnEqsDimensional}
\end{equation}
This form of the equations is convenient for the study of the stationary
deflagration front as it allows one to immediately obtain the first intergal
of the heat-conduction equation.

Alternatively one can use
\begin{equation}
K=\int_{T_{0}}^{T}k(T^{\prime })dT^{\prime }  \label{KDef}
\end{equation}
as the temperature variable, with $T_{0}$ being a reference temperature.
With this choice$,$ the first of Eqs.\ (\ref{TnEqsDimensional}) takes the
form
\begin{equation}
\frac{1}{\kappa }\frac{\partial K}{\partial t}=\nabla ^{2}K-\Delta E\frac{%
\partial n_{-}}{\partial t},  \label{KEq}
\end{equation}
while in the second equation one should use $T=T(K)$ in the expression for $%
\Gamma .$ This form of equations is convenient for the study of the
deflagration threshold in the case when the temperature along the boundary
of the crystal varies.

\section{Thermal runaway}

\label{Sec-Runaway}

Deflagration begins with a thermal runaway in a part of the sample that has
a lower barrier $U$ or a higher temperature $T$ than the surrounding area.
Thermal runaway needs some time to develop. We call it the ignition time $%
\tau _{\mathrm{ig}}$. The shortest ignition time is achieved if the heat
released by the relaxation remains in the sample and does not escape through
its boundaries. We denote this ignition time $\tau _{\mathrm{ig}}^{(\infty
)},$ as it is related to the ignituin in the infinite sample, see below. If
the rate of heat transfer out of the sample is sufficiently high, the
ignition does not occur. This explains why small crystals do not exhibit
magnetic avalanches. In this section we derive the expression for the
ignition rate $\Gamma _{\mathrm{ig}}^{(\infty )}\equiv 1/\tau _{\mathrm{ig}%
}^{(\infty )}$ that plays fundamental role in subsequent considerations and
justifies the validity of the explosive approximation $n_{-}\Rightarrow
n_{-,i}$ ($n_{-,i}$ is the initial value of $n_{-}$) that will be used below.

For an infinite and/or thermally insulated sample one can drop the diffusion
term in the first of Eqs.\ (\ref{TnEqsDimensional}). This yields
\begin{equation}
\frac{\partial T}{\partial t}=\frac{\Delta E}{C_{\mathrm{ph}}}\Gamma
(T)n_{-}.  \label{TEqIgnition}
\end{equation}
The initial conditions are $T=T_{0}$ and $n_{-}=n_{-,i}.$ Using Eq.\ (\ref
{GammaRes}) in the strong-bias case, it is convenient to introduce the
reduced temperature deviation and the reduced population of the metastable
well
\begin{equation}
\theta \equiv W_{0}\frac{T-T_{0}}{T_{0}},\qquad \tilde{n}\equiv \frac{n_{-}}{%
n_{-,i}}  \label{thetaDef}
\end{equation}
with $W_{0}$ defined by Eq.\ (\ref{WDef}) with $T=T_{0}.$ Linearization of
the argument of $\Gamma (T)$ on $\theta $ leads to the system of equations
\begin{equation}
\frac{\partial \theta }{\partial \tau }=e^{\theta }\tilde{n},\qquad \nu _{0}%
\frac{\partial \tilde{n}}{\partial \tau }=-e^{\theta }\tilde{n}
\label{thetatnEqs}
\end{equation}
where $\tau \equiv \Gamma _{\mathrm{ig}}^{(\infty )}t$ is the reduced time.
The ignition rate is given by
\begin{equation}
\Gamma _{\mathrm{ig}}^{(\infty )}=\nu _{0}\Gamma (T_{0}),  \label{Gammaiginf}
\end{equation}
where
\begin{equation}
\qquad \nu _{0}\equiv W_{0}\frac{n_{-,i}\Delta E}{C_{\mathrm{ph,0}}T_{0}}%
=W_{0}^{2}\frac{n_{-,i}\Delta E}{U}\frac{k_{B}}{C_{\mathrm{ph,0}}}
\label{nuDef}
\end{equation}
and $C_{\mathrm{ph,0}}$ is the phonon heat capacity at $T=T_{0}.$ Note that
in cases of practical interest $\nu _{0}$ is a large parameter since it
contains $W_{0}^{2},$ whereas $\Delta E/U$ is typically of order one. The
factor $k_{B}/C_{\mathrm{ph,0}}$ is also large at small temperature $T_{0},$
see Eq.\ (\ref{Cph3d}). Since $\nu _{0}\gg 1,$ the evolution of $n_{-}$ is
much slower than that of $\theta ,$ so that one can approximately replace $%
n_{-}$ with $n_{-,i}$ to study ignition. This is what we call the explosive
approximation or the explosive limit. (Indeed, for explosives the parameter $%
\nu _{0}$ is very large, so that the problem of their stability can be
considered without taking into account that a small part of the explosive
has already burned and the heat release has been reduced because of this.)

Setting $n_{-}\Rightarrow n_{-,i}$ in the first of Eqs.\ (\ref{thetatnEqs})
one obtains an isolated equation that has the solution $\theta (\tau )=-\ln
(1-\tau )$ reaching infinity exactly at $\tau =1.$ In real units, it
corresponds to $t=\tau _{\mathrm{ig}}^{(\infty )}.$ Of course, soon after
the ignition the deviation of $T$ from $T_{0}$ becomes large, thus the
linearized equation $\partial _{\tau }\theta =e^{\theta }$, as well as the
replacement $n_{-}\Rightarrow n_{-,i},$ becomes invalid. Nevertheless, the
dominant contribution into the ignition time comes from the time range $%
\theta \sim 1$ where still $T-T_{0}\ll T_{0}$ and the equation $\partial
_{\tau }\theta =e^{\theta }$ is valid.

\begin{figure}[t]
\unitlength1cm
\begin{picture}(11,5.5)
\centerline{\psfig{file=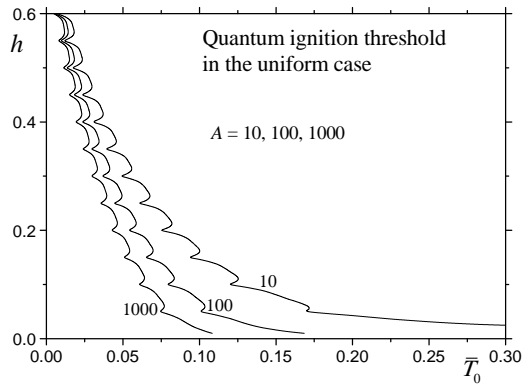,angle=-90,width=8cm}}
\end{picture}
\caption{The dependence of the magnetic field that ignites deflagration on
the temperature of the crystal of Mn$_{12}.$ Dimensionless parameters are
given by Eqs.\ (\ref{U0Def}) and (\ref{TbarADef}). }
\label{Fig-QDT}
\end{figure}

The low-temperature explosive approximation $n_{-}\Rightarrow n_{-,i}$
introduced above drastically simplifies the problem of the ignition of
deflagration and allows one to understand it in simple terms. With $%
n_{-}=n_{-,i},$ the temperature of the sample is the only relevant variable,
and its dynamics is determined by the competition of the two terms. One of
them is the heat release due to the relaxation that is strongly nonlinear in
temperature. The other one is the heat loss due to the heat conduction that
is linear on temperature but contains spatial derivatives. There are two
scenarious if one starts with the sample having an uniform temperature $%
T=T_{0}$ that coincides with the constant temperature of the sample
boundaries. In the first scenario, the temperature increases because of the
heat release, typically with a maximum at the center of the sample, until a
sufficient temperature gradient develops that provides the balance between
the heat release and heat loss through the boundaries. The resulting state
is the stationary state of the system. In the second scenario, the heat loss
through the boundaries is insufficient to balance the increase of the
heat-release due to the rise of temperature. This happens, in particular, if
the sample is sufficiently large. In this case there is no stationary state,
and the temperature growth, slow at the beginning, leads to a thermal
runaway. Changing one of the parameters (sample size, energy barrier,
temperature at the boundaries, initial magnetization) one can reach the
situation in which the stationary state disappears and the runaway begins.
We call it the ignition threshold. Below we present analytical and numerical
results for the ignition threshold in different cases in the explosive
limit. We will see that the ignition is mainly controlled by the parameter
\begin{equation}
\delta \equiv \frac{\Gamma _{\mathrm{ig}}^{(\infty )}}{\Gamma _{\kappa }}%
=\left( \frac{R}{l_{0}}\right) ^{2}=\frac{R^{2}U\Delta En_{-,i}\,\Gamma
(T_{0})}{2k_{0}k_{B}T_{0}^{2}},  \label{deltaDef}
\end{equation}
where $R$ is a typical shortest distance from the center of the sample to
its boundary and $k_{0}$ is thermal conductivity $k$ at $T=T_{0}.$ For a $1d$
sample (i.e., a cylinder thermally insulated along its side) one has $R=L/2,$
where $\ L$ is the sample thickness, whereas for cylindrical and spherical
samples $R$ is the radius. $\Gamma _{\mathrm{ig}}^{(\infty )}$ is given by
Eq.\ (\ref{Gammaiginf}) while
\begin{equation}
\Gamma _{\kappa }=2\kappa _{0}/R^{2}  \label{GammakappaDef}
\end{equation}
is the rate of thermal equilibration within the sample at $T=T_{0}.$ In Eq.\ (%
\ref{deltaDef}) $l_{0}$ is the characteristic thermal length
\begin{equation}
l_{0}=\sqrt{2\kappa _{0}/\Gamma _{\mathrm{ig}}^{(\infty )}}  \label{l0Def}
\end{equation}
at $T=T_{0}.$

We will see that in the simplest case of the uniform energy barrier and
constant temperature $T_{0}$ maintained at the boundaries (uniform
conditions) the ignition threshold corresponds to $\delta =\delta _{c}\sim
1. $ The exact value $\delta _{c}$ depends on the geometry of the sample. In
particular, in one dimension $\delta _{c}\simeq 0.439.$ This allows one to
obtain a relation between the temperature $T_{0}$ and the barrier $U$\ \ at
the ignition threshold, that also depends on the sample size $R$ and other
parameters. In terms of dimensionless parameters
\begin{equation}
\bar{T}_{0}\equiv \frac{k_{B}T_{0}}{U_{0}},\qquad A\equiv \frac{R^{2}\Gamma
_{0}n_{-,i}}{2k(\bar{T}_{0})/k_{B}}  \label{TbarADef}
\end{equation}
[see Eqs.\ (\ref{U0Def}) and (\ref{GammaRes})] one can write Eq.\ (\ref
{deltaDef}) in the form
\begin{equation}
\delta =A\frac{4hu(h)}{\bar{T}_{0}^{2}}\exp \left[ -\frac{u(h)}{\bar{T}_{0}}%
\right] ,  \label{deltared}
\end{equation}
where $u(h)$ is given by $u(h)=(1-h)^{2}$ for the classical model and by Eq.\
(\ref{UQDef}) with account of spin tunneling$.$ Resolving the threshold
equation $\delta =\delta _{c}$ requires the knowledge of the temperature
dependence of thermal conductivity $k.$ It turns out, however (see below)
that the exact form of $k(\bar{T}_{0})$ given by Eq.\ (\ref{kPowerLaw}) is
not essential. The results for $k(T)=\mathrm{const}$ and thus $A=\mathrm{%
const}$ are shown in Fig.\ \ref{Fig-QDT} for three different values of $A.$
One can see that increasing $A$ (say, due to the increasing of the sample
size $R)$ leads to the decrease of the critical values of $\bar{T}_{0}$ and $%
h$. In the realistic case of the large Arrhenius exponent in Eq.\ (\ref
{deltared}), the dependence on $A$ and thus on $k$ is logarithmic.

\begin{figure}[t]
\unitlength1cm
\begin{picture}(11,5.5)
\centerline{\psfig{file=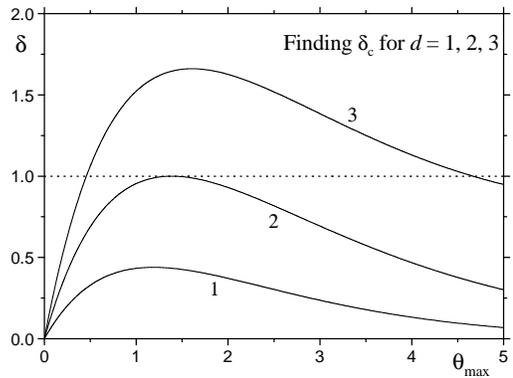,angle=-90,width=8cm}}
\end{picture}
\caption{Plot of $\protect\delta (\protect\theta _{\max })$ for $d=1,2,3$
that allows one to obtain the ignition threshold from the maximum of these curves.}
\label{Fig-deltacd123}
\end{figure}

\section{Ignition threshold in molecular magnets}

\label{Sec-threshold}

\subsection{Ignition threshold under uniform conditions}

\subsubsection{Ignition threshold in one dimension}

Consider a slab of thickness $L=2R$. We will see that at $W\gg 1,$ if the
constant temperature $T_{0}$ is maintained at the boundaries of the slab$,$
the solution for $T$ at the ignition threshold only slightly deviates from $%
T_{0}.$ In this case one can linearize the problem around $T_{0}$ using the
temperature deviation $\theta $ of Eq.\ (\ref{thetaDef}) and set $%
k\Rightarrow k_{0}.$ With $\partial T/\partial t=0$ in the stationary case
one obtains the equation
\begin{equation}
\frac{d^{2}\theta }{dx^{2}}+2\delta e^{\theta }=0,\qquad \theta (\pm 1)=0,
\label{thetaEq1}
\end{equation}
where the reduced space variable $x$ is normalized by $R$. The first
integral of this equation is
\begin{equation}
\left( \frac{d\theta }{dx}\right) ^{2}+4\delta \left( e^{\theta }-e^{\theta
_{\max }}\right) =0  \label{thetaEq}
\end{equation}
where $\theta _{\max }$ is the integration constant that equals to the
maximal value of $\theta $ achieved in the middle of the sample. Integrating
Eq.\ (\ref{thetaEq}) one obtains
\begin{equation}
\theta (x)=\theta _{\max }-2\ln \cosh \left( \sqrt{\delta e^{\theta _{\max }}%
}x\right) ,  \label{thetex1dRes}
\end{equation}
where the value of $\theta _{\max }$ follows from the boundary conditions $%
\theta (\pm 1)=0$. To find the ignition threshold, one can solve this
equation for $\delta $:
\begin{equation}
\delta =e^{-\theta _{\max }}\ln ^{2}\left[ e^{\theta _{\max }/2}+\sqrt{%
e^{\theta _{\max }}-1}\right] .  \label{deltathetamax1d}
\end{equation}
The dependence $\delta (\theta _{\max })$ is shown in Fig.\ \ref
{Fig-deltacd123}. It has a maximum at $\theta _{\max }=\theta _{\max
,c}=1.18684.$ The maximal value of $\delta $%
\begin{equation}
\delta _{c}=0.439229,  \label{deltac1d}
\end{equation}
corresponds to the ignition threshold. Indeed, for $\delta <\delta _{c}$
there are two solutions for $\theta _{\max }$, and the smallest of the two
corresponds to the stationary solution of the heat-conduction equation. For $%
\delta >\delta _{c}$ the stationary solution disappears.

\subsubsection{Ignition threshold in two and three dimensions}

For a cylindrical ($d=2$) and spherical ($d=3$) samples the generalization
of Eq.\ (\ref{thetaEq1}) is
\begin{equation}
\frac{d^{2}\theta }{dr^{2}}+\frac{d-1}{r}\frac{d\theta }{dr}+2\delta
e^{\theta }=0  \label{thetaEqd}
\end{equation}
with $r$ normalized by $R$ and with the boundary conditions $\theta ^{\prime
}(0)=0$ and $\theta (1)=0.$ For $d=2$ the exact solution of Eq.\ (\ref
{thetaEqd}) is
\begin{equation}
\theta (r)=2\ln \frac{2}{1+\sqrt{1-\delta }+\left( 1-\sqrt{1-\delta }\right)
r^{2}}.  \label{thetar2d}
\end{equation}
Its maximal value
\begin{equation}
\theta _{\max }=2\ln \frac{2}{1+\sqrt{1-\delta }}  \label{thetamax2d}
\end{equation}
is achieved at $r=0$. The ignition threshold can be found by the same method
as in $1d.$ Resolving this equation for $\delta $ one obtains
\begin{equation}
\delta =4e^{-\theta _{\max }}\left( e^{\theta _{\max }/2}-1\right) .
\label{deltathetamax2d}
\end{equation}
This function has a maximum at $\theta _{\max }=\theta _{\max \mathrm{,c}%
}=2\ln 2\simeq 1.\,\allowbreak 386\,3,$ and the corresponding critical value
of $\delta $ is
\begin{equation}
\delta _{c}=1.  \label{deltac2d}
\end{equation}

For $d=3$ we are unable to find the solution of Eq.\ (\ref{thetaEqd}) in
terms of known functions. Numerical solution for the ignition threshold in $%
3d$ consists of the following steps: (i) One solves Eq.\ (\ref{thetaEqd})
with the boundary conditions $\theta ^{\prime }(0)=0$ and $\theta (0)=\theta
_{\max }\ $and $\delta $ as a free parameter; (ii) One finds $\delta $ as a
function of $\theta _{\max }$ from the boundary condition $\theta (1)=0;$
(iii) One finds critical parameters from the maximum of $\delta (\theta
_{\max }).$ Our results for $d=1,2,3$ are listed below
\begin{equation}
\begin{tabular}{lllll}
$d\quad $ & $\delta _{c}$ & $\delta _{c}/d$ & $1/\delta _{c}$ & $\theta
_{\max \mathrm{,c}}$ \\
1 & 0.4392$\quad $ & 0.4392$\quad $ & 2.277 & 1.187 \\
2 & 1 & 0.5 & 1 & 1.386 \\
3 & 1.661 & 0.5537 & 0.6020$\quad $ & 1.607
\end{tabular}
\label{critparsd123}
\end{equation}
One can see that approximately $\delta _{c}\varpropto d.$ The curves $\delta
(\theta _{\max })$ are plotted in Fig.\ \ref{Fig-deltacd123}.

\begin{figure}[t]
\unitlength1cm
\begin{picture}(11,5.5)
\centerline{\psfig{file=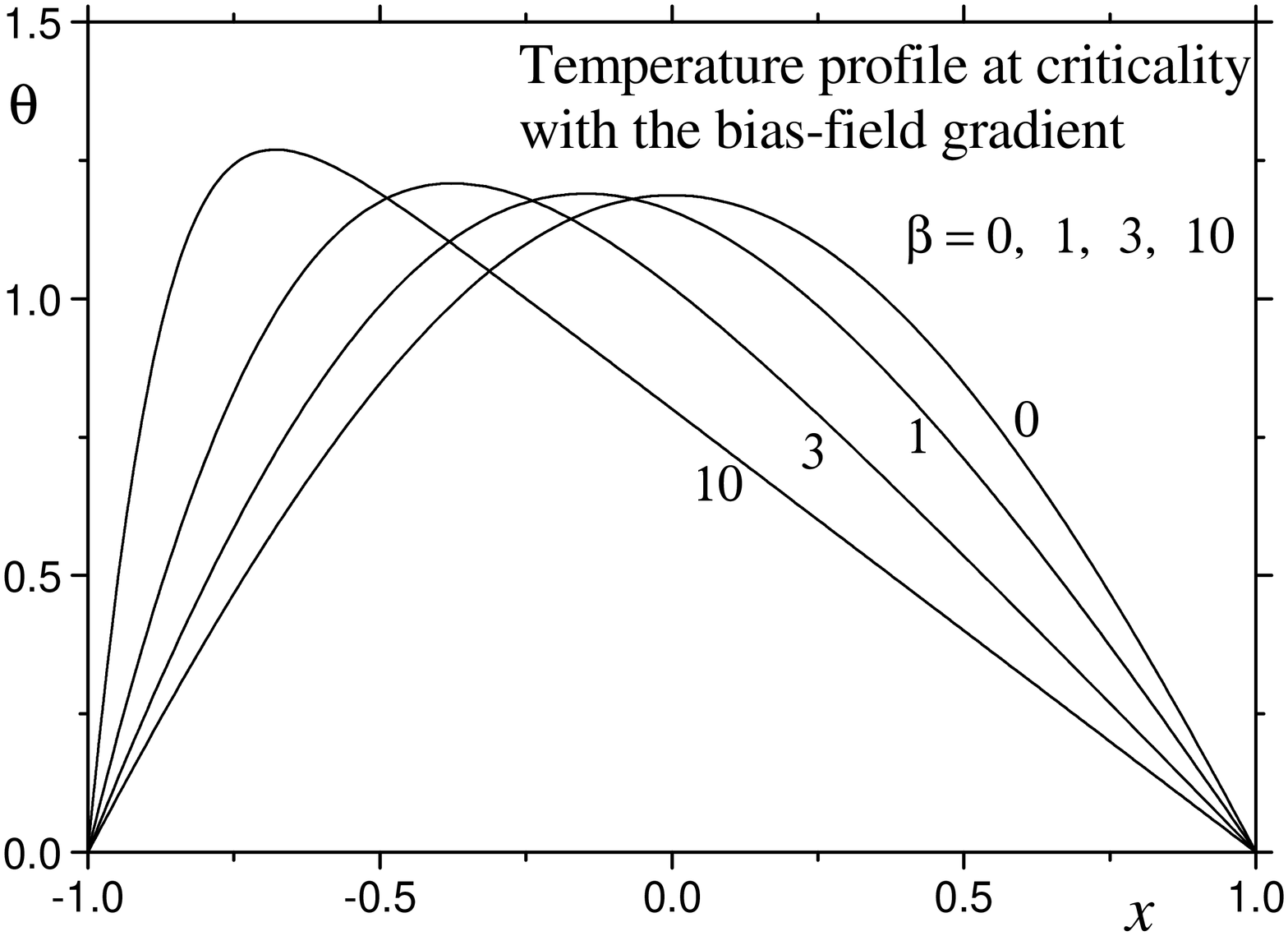,angle=-90,width=8cm}}
\end{picture}
\begin{picture}(11,5.5)
\centerline{\psfig{file=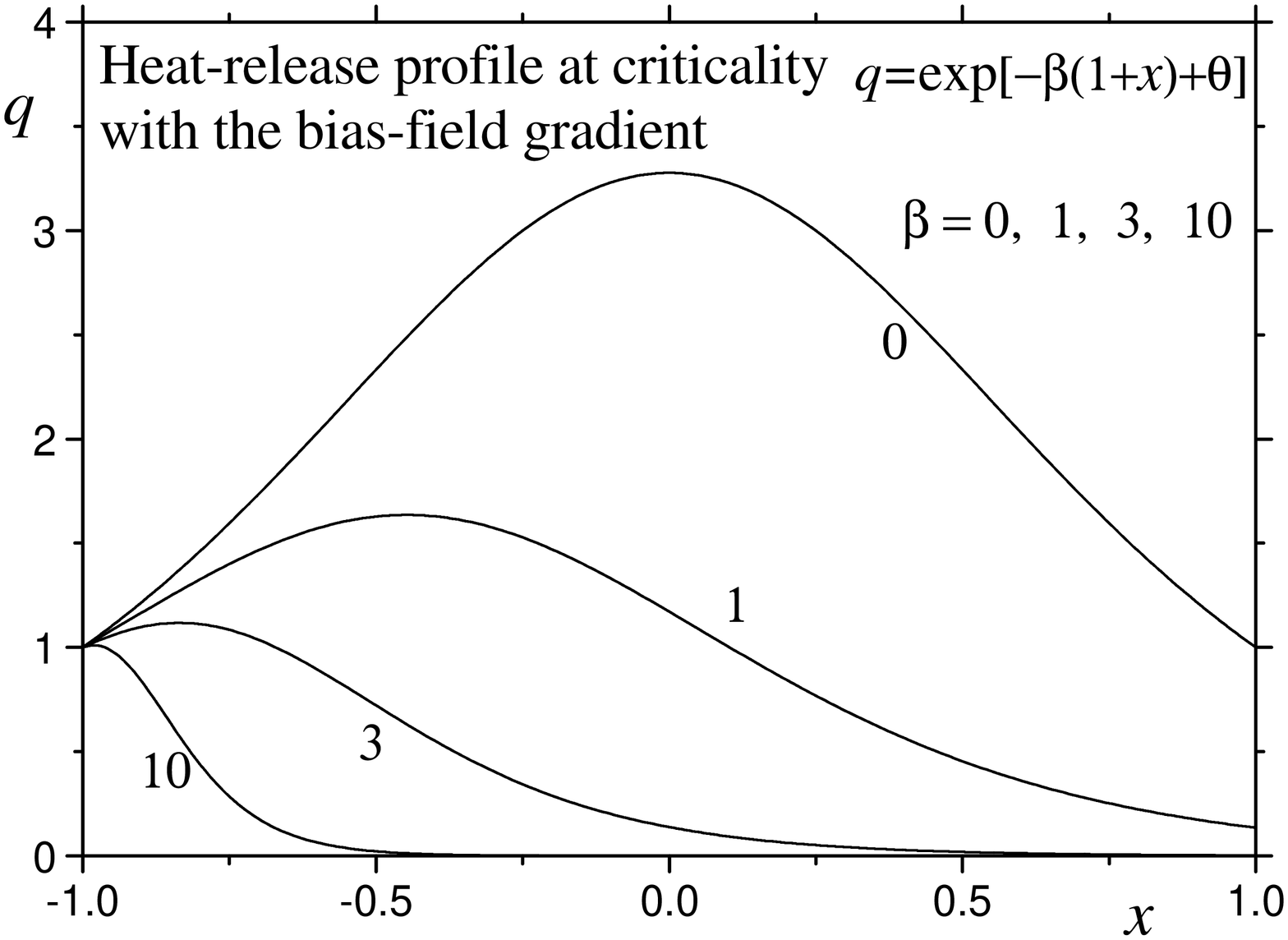,angle=-90,width=8cm}}
\end{picture}
\caption{Numerical results for temperature and heat-release profiles.}
\label{Fig-fieldgradient}
\end{figure}
\begin{figure}[t]
\unitlength1cm
\begin{picture}(11,5.5)
\centerline{\psfig{file=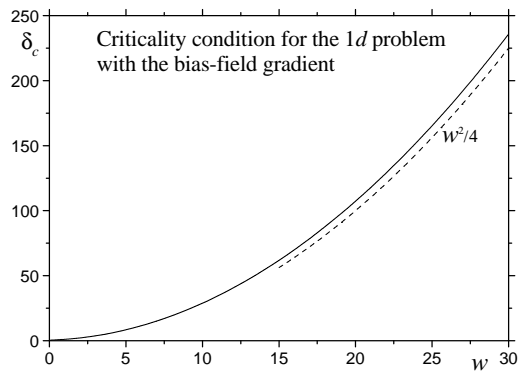,angle=-90,width=8cm}}
\end{picture}
\caption{Ignition threshold in the model with the bias-field gradient.}
\label{Fig-delta_c-fieldgradient}
\end{figure}

\subsection{Ignition threshold in the presence of field gradient}

Consider a one-dimensional problem of ignition with temperature at both ends
maintained at $T_{0}$ and the barrier $U$ varying in space due to the
gradient of the bias field. Although the relative variation of $U$ is small,
the effect can be large for large Atthenius factors $W$ as its variation $%
\delta W$ can be large. Assuming that the barrier is the lowest at the left
end of the sample and the field gradient is small and constant, one can
write
\begin{equation}
\delta W\cong w(1+x)-\theta .  \label{deltaBDef}
\end{equation}
The equation for the stationary temperature profile becomes
\begin{equation}
\frac{d^{2}\theta }{dx^{2}}+2\delta e^{-w(1+x)+\theta }=0,\qquad \theta (\pm
1)=0.  \label{thetaEqbeta}
\end{equation}
For $w\sim 1,$ Eq.\ (\ref{thetaEq1}) can only be solved numerically. Here,
instead of plotting $\delta $ vs $\theta _{\max },$ it is more convenient to
plot $\delta $ vs $\theta ^{\prime }(-1)$. Numerical results for temperature
profile $\theta (x)$ and heat-release profile $q(x)\varpropto \exp
[-w(1+x)+\theta ]$ \ at the ignition threshold are shown in Fig.\ \ref
{Fig-fieldgradient}. As the field gradient goes up, the maxima of these
curves shift towards the end of the sample where the barrier is lower. The
threshold condition $\delta _{c}(w)$ is shown in Fig.\ \ref
{Fig-delta_c-fieldgradient}. The value of $\delta _{c}$ increases with $w$
since the favorable condition for burning is realized in a more and more
narrow region at the left end of the sample, and there is an increasing heat
flow out of this region in both directions.

In the case of $w\gg 1$ the ignition occurs very close to the left end, $%
x=-1,$ and the heat release proportional to $\exp \left[ -w(1+x)+\theta %
\right] $ is very close to zero except in the vicinity of the left end. The
temperature profile for $w\gg 1$ consists of two regions: Very close to the
left end the temperature rises sharply to the maximal temperature $\theta
_{\max }$ and then goes linearly down to zero at the right end. Thus for $%
w\gg 1$ one can introduce a new variable $u\equiv w(1+x),$ a new function $%
\vartheta =\theta -u,$ and replace Eq.\ (\ref{thetaEqbeta}) by
\begin{equation}
\frac{d^{2}\vartheta }{du^{2}}+2\tilde{\delta}e^{\vartheta }=0,\qquad \tilde{%
\delta}\equiv \frac{\delta }{w^{2}}  \label{varthetaEq}
\end{equation}
with the boundary conditions $\vartheta (0)=0,$ $\vartheta ^{\prime }(\infty
)=-1.$ This equation is similar to Eq.\ (\ref{thetaEq1}) and its solution
reads
\begin{equation}
\vartheta (u)=\vartheta _{\max }-\ln \cosh ^{2}\left( \sqrt{\tilde{\delta}%
e^{\vartheta _{\max }}}\left( u-u_{\max }\right) \right) ,
\end{equation}
where $\vartheta _{\max }$ and $u_{\max }$ are integration constants. From
the boundary condition $\vartheta ^{\prime }(\infty )=-1$ one obtains $2%
\sqrt{\tilde{\delta}e^{\vartheta _{\max }}}=1$ and thus
\begin{equation}
\vartheta _{\max }=\ln \frac{1}{4\tilde{\delta}}.  \label{ttetaMax}
\end{equation}
Then the other boundary condition, $\vartheta (0)=0,$ gives
\begin{equation}
u_{\max }=2\,\mathrm{arccosh}\frac{1}{2\sqrt{\tilde{\delta}}}.
\label{uMaxRes}
\end{equation}
Since $\vartheta _{\max }\geq 0,$ the ignition threshold is defined by
\begin{equation}
\tilde{\delta}_{c}=\frac{1}{4},\qquad \delta _{c}=\frac{w^{2}}{4}.
\label{deltacbeta}
\end{equation}
For $\delta =\delta _{c}$ one has $u_{\max }=0$ $,$ that is, the ignition
occurs at the left boundary. Corrections to Eq.\ (\ref{varthetaEq}) move the
maximal-temperature point a little away from the edge of the sample.
Comparing this situation with the uniform case with the same barrier as the
barrier at the left end (the minimal barrier), one can see that in the case
of $w\gg 1$ the ignition requires a much higher temperature $T_{0}$.

The constant $w$ in Eq.\ (\ref{deltaBDef}) can be expressed via the gradient
of the bias field with the help of Eq.\ (\ref{UDef}):
\begin{equation}
w=\frac{dW}{dx}=W_{0}(1-h)\left| \Delta h\right| ,  \label{wRes}
\end{equation}
where $\Delta h$ is the change of $h$ across the sample. In terms of the
energy bias $\Delta E$ defined by Eq.\ (\ref{DeltaEDef}) one has
\begin{equation}
w=\frac{1-h}{4}\frac{\Delta \left( \Delta E\right) }{k_{B}T_{0}},
\label{wDeltaDeltaE}
\end{equation}
where $\Delta \left( \Delta E\right) $ is the change of the energy bias
across the sample. Note that at a very low temperature $T_{0},$ the
condition $w\gg 1$ does not necessarily require a large field gradient.
Because of the field gradient, the Arrhenius factor $W$ increases by 1 at
the characteristic distance
\begin{equation}
l_{H}\equiv \frac{2R}{w}=\frac{L}{w}=\frac{L}{W_{0}(1-h)\left| \Delta
h\right| }  \label{lhDef}
\end{equation}
from the end. It is the width of the ignition region near the low-barrier
end. In the case of $w\gg 1,$ that is $l_{H}\ll L,$ the far end with the
highest barrier becomes irrelevant for the ignition, and the ignition
threshold $\delta =\delta _{c}$ that follows from Eqs.\ (\ref{deltaDef}) and (%
\ref{deltacbeta}) becomes
\begin{equation}
l_{H}=l_{0}.  \label{CriticalityCondbeta}
\end{equation}
For a cylinder of radius $R$ with the bias field linearly changing along its
symmetry axis, this condition holds if $l_{H}\ll R,$ so that the heat flows
along the cylinder axis $z$ away from the face with the lowest barrier
rather than towards the side walls of the cylinder. The problem then becomes
one dimensional.

\begin{figure}[t]
\unitlength1cm
\begin{picture}(11,5.5)
\centerline{\psfig{file=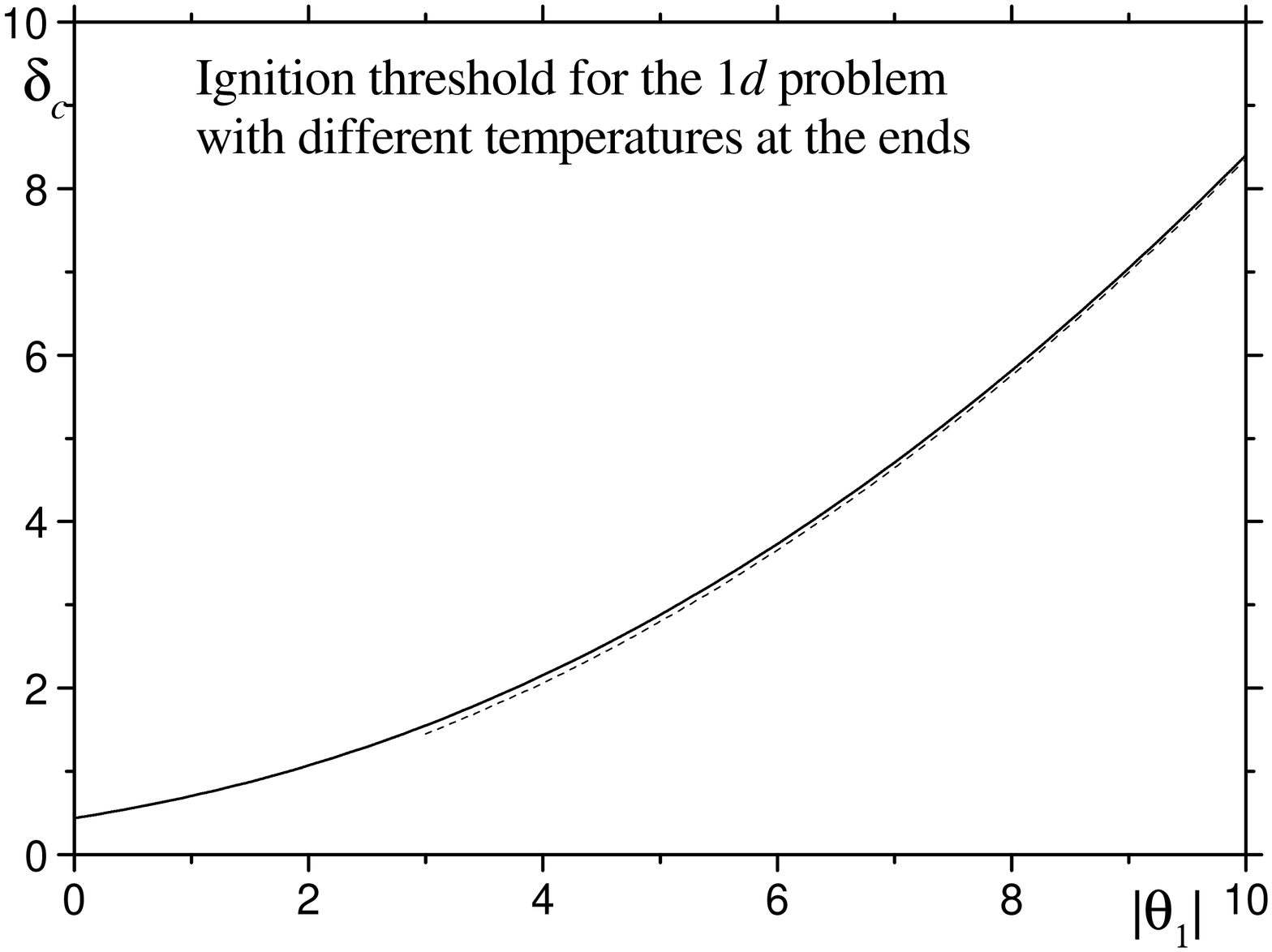,angle=-90,width=8cm}}
\end{picture}
\begin{picture}(11,5.5)
\centerline{\psfig{file=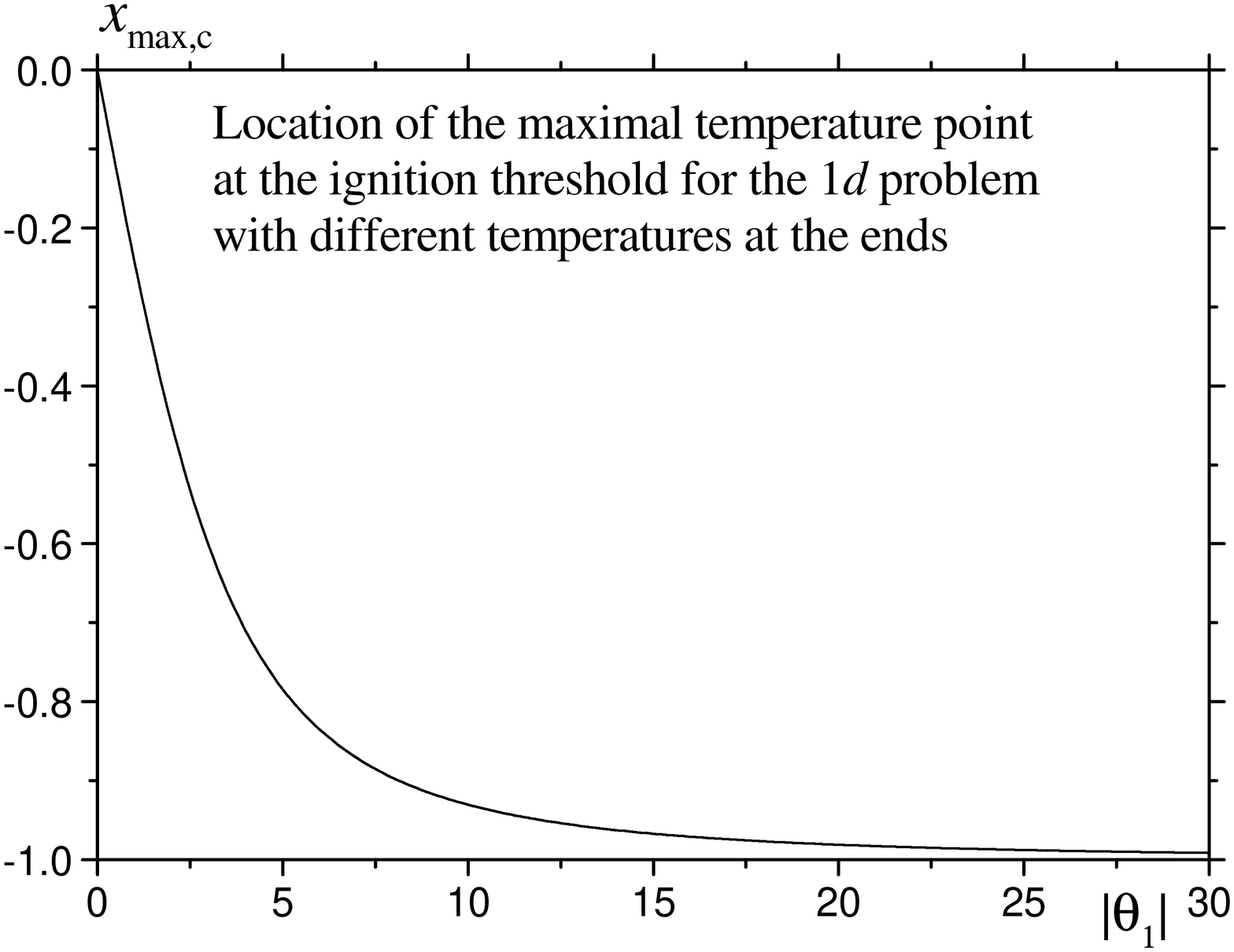,angle=-90,width=8cm}}
\end{picture}
\caption{Numerical results for $\protect\delta _{c}$ (a) and $x_{\max ,%
\mathrm{c}}$ (b) vs the temperature-bias parameter $\protect\theta _{1}$ for
the $1d$ model with different temperatures at the ends. $x=0$ corresponds to
the center of the sample. The dashed line in (a) is the asymptote $\protect%
\delta _{c}\cong \left( |\protect\theta _{1}|+2\ln 2\right) ^{2}/16+1/4$ at $%
|\protect\theta _{1}|\gg 1.$}
\label{Fig-theta1-dep}
\end{figure}

\subsection{Ignition threshold in the presence of temperature gradient}

\label{Sec-TempGradient}

While it is experimentally difficult to create a large gradient of the
bias-field over the length of a small crystal, it is relatively easy to
break the symmetry of the system by creating a large temperature gradient.
This can be done by, e.g., maintaining temperature $T_{0}$ at the left end
and having $T_{1}<T_{0}$ at the right end of a $1d$ sample. If these two
temperatures differ essentially, one has to take into account the
temperature dependence of thermal conductivity $k(T)$ that is strong at low $%
T,$ see Eq.\ (\ref{kPowerLaw}). In this case it is more convenient to use $K$
defined by Eq.\ (\ref{KDef}) instead of $T.$ As the ignition occurs closer to
the hot left end, it is convenient to choose $T_{0}$ as the reference
temperature and introduce
\begin{equation}
\theta \equiv W_{0}\frac{K}{k_{0}T_{0}}\,  \label{thetaKDef}
\end{equation}
that generalizes Eq.\ (\ref{thetaDef}), with $k_{0}\equiv k(T_{0}).$ The
relaxation rate can be expanded similarly to the above,
\begin{equation}
\Gamma (T(K))\cong \Gamma (T_{0})\exp \left( W_{0}\frac{\delta T}{T_{0}}%
\right) \cong \Gamma (T_{0})e^{\theta }.  \label{GammaThetaExp}
\end{equation}
The applicability of this expansion requires $|\delta T|/T_{0}\ll 1.$
However, in the case of $W_{0}\gg 1$ this expansion practically works in the
whole range of $\delta T<0$ since the burning rate $\Gamma $ becomes
negligibly small long before the condition $|\delta T|/T_{0}\ll 1$ is
violated. In the stationary state $\theta (x)$ satisfies the same Eq.\ (\ref
{thetaEq1}) but with the boundary conditions
\begin{equation}
\theta (-1)=0,\qquad \theta (1)\equiv \theta _{1}=W_{0}\frac{K(T_{1})}{%
k_{0}T_{0}}\leq 0.  \label{thetaBC}
\end{equation}
In the absense of the heat release due to burning the solution for $\theta $
would be a linear function, $\theta (x)=(1+x)\theta _{1}/2.$ An estimate for
the width $l_{T}$ of the region near the hot end where the ignition occurs
can be obtained by setting $\theta (x)\sim -1$. In real units ignition
occurs at the distance of order
\begin{equation}
l_{T}=4R/\left| \theta _{1}\right| ,  \label{lTDef}
\end{equation}
where the numerical factor 4 anticipates Eq.\ (\ref{deltacT0T1}). For the
very cold right end, $\left| \theta _{1}\right| $ $\gg 1,$ one has $l_{T}\ll
R$. The first integral of Eq.\ (\ref{thetaDef}) is Eq.\ (\ref{thetaEq}) in
which the maximum $\theta =\theta _{\max }$ can be achieved at some $x_{\max
}$ shifted from the central point $x=0.$ The solution for $\theta (x)$ reads
\begin{equation}
\theta (x)=\theta _{\max }-2\ln \cosh \left[ \sqrt{\delta e^{\theta _{\max }}%
}\left( x-x_{\max }\right) \right]  \label{thetaxTbias}
\end{equation}
which is the generalization Eq.\ (\ref{thetex1dRes}). Elimination of $x_{\max
}$ using the boundary conditions and solving for $\delta $ yields
\begin{eqnarray}
\delta &=&\frac{1}{4}e^{-\theta _{\max }}\ln ^{2}\left[ \left( e^{\theta
_{\max }/2}+\sqrt{e^{\theta _{\max }}-1}\right) \right.  \notag \\
&&\times \left. \left( e^{\left( \theta _{\max }-\theta _{1}\right) /2}+%
\sqrt{e^{\theta _{\max }-\theta _{1}}-1}\right) \right] ,
\label{deltaviathetamax}
\end{eqnarray}
c.f. Eq.\ (\ref{deltathetamax1d}). Computing the maximum of this function on $%
\theta _{\max }$ allows to determine $\delta _{c}$ for any value of $\theta
_{1}.$

If the temperature at the right end is low, $-\theta _{1}$ is a large
positive value, so that Eq.\ (\ref{deltaviathetamax}) simplifies. In this
case at the ignition threshold one has $\theta _{\max }\ll 1$, so that Eq.\ (%
\ref{deltaviathetamax}) becomes
\begin{equation}
\delta \cong \frac{\theta _{1}^{2}}{16}e^{-\theta _{\max }}\left( 1+\frac{%
4\ln 2}{|\theta _{1}|}+\frac{4\sqrt{\theta _{\max }}}{|\theta _{1}|}\right) .
\end{equation}
The maximum of the rhs is attained at $\theta _{\max }=\theta _{\max
,c}\cong 4/\theta _{1}^{2}\ll 1.$ Thus the ignition threshold is defined by
\begin{equation}
\delta _{c}\cong \left( \frac{|\theta _{1}|+2\ln 2}{4}\right) ^{2}+\frac{1}{4%
}\cong \left( \frac{W_{0}}{4}\frac{K(T_{1})}{k_{0}T_{0}}\right) ^{2}.
\end{equation}
The maximal-temperature point at the ignition threshold is
\begin{equation}
x_{\max ,\mathrm{c}}\cong -1+8/\theta _{1}^{2},  \label{xmaxc}
\end{equation}
which is close to the left end. One can see that for $|\theta _{1}|\gg 1$
the value of $\delta _{c}$ is large, so that much larger $L$ is needed to
reach the threshold for the same temperature $T_{0}$, as compared to the
uniform case. The ignition threshold $\delta =\delta _{c}$ is equivalent to
\begin{equation}
l_{T}=l_{0},  \label{deltacT0T1}
\end{equation}
where $l_{T}$ is given by Eq.\ (\ref{lTDef}). This result is similar to Eq.\ (%
\ref{CriticalityCondbeta}). The numerically obtained dependences $\delta
_{c}(|\theta _{1}|)$ and $x_{\max ,\mathrm{c}}(|\theta _{1}|)$ are shown in
Fig.\ \ref{Fig-theta1-dep}.

The remaining task is to relate the temperature-bias parameter $\theta _{1}$
to the temperatures at the ends, $T_{0}$ and $T_{1}.$ At a constant thermal
conductivity $k$ one obtains
\begin{equation}
\theta _{1}=-W_{0}\left( 1-\frac{T_{1}}{T_{0}}\right)
\end{equation}
that tends to $-W_{0}$ in the limit $T_{1}\rightarrow 0.$ Since $W_{0}\gg 1,$
there is a strong suppression of the ignition by the cold end. The effect is
even stronger for the power-law dependence of $k$ given by Eq.\ (\ref
{kPowerLaw}). Parametrization
\begin{equation}
k(T)=k_{0}\left( \frac{T}{T_{0}}\right) ^{-\gamma }
\end{equation}
and integration in Eq.\ (\ref{KDef}), with the lower limit being $T_{0},$
yields
\begin{equation}
K(T)=-\frac{k_{0}T_{0}}{\gamma -1}\left[ \left( \frac{T_{0}}{T}\right)
^{\gamma -1}-1\right] .
\end{equation}
Now one obtains
\begin{equation}
\theta _{1}=-\frac{W_{0}}{\gamma -1}\left[ \left( \frac{T_{0}}{T_{1}}\right)
^{\gamma -1}-1\right]
\end{equation}
and for $|\theta _{1}|\gg 1$
\begin{equation}
\delta _{c}\cong \left\{ \frac{W_{0}}{4\left( \gamma -1\right) }\left[
\left( \frac{T_{0}}{T_{1}}\right) ^{\gamma -1}-1\right] \right\} ^{2}.
\end{equation}
In the realistic case given by Eq.\ (\ref{kPowerLaw}) one has $\gamma -1=1/3.$
That is, if $T_{1}$ goes to zero, $\theta _{1}$ becomes infinite negative
and $\delta _{c}$ diverges. This means that for a sample of any size one can
suppress the ignition by making the temperature of the cold end very close
to zero. This is a consequence of the divergence of thermal conductivity at $%
T=0.$

\section{Rate of ignition of deflagration in molecular magnets}

\label{Sec-Ignition-rate}

Above the ignition threshold, $\delta >\delta _{c},$ the heat loss via heat
conduction cannot compensate the heat release due to burning and there is no
stationary solution for the temperature$.$ The temperature growth leads to a
thermal runaway after the ignition time $\tau _{\mathrm{ig}}$, followed by
the deflagration. At first we investigate the ignition time within the
explosive approximation, $n_{-}\Rightarrow n_{-,i},$ in terms of $\theta $
defined by Eqs.\ (\ref{thetaDef}) or (\ref{thetaKDef}). Then we study
deviations from the explosive approximation using a more general system of
equations containing both $\theta $ and $n_{-}.$ The initial condition in
all examples considered in this section is thermal equilibrium reached in
the absence of relaxation, for instance, a uniform temperature throughout
the sample. This is the most transparent case theoretically but it may be
difficult to realize in experiment if the ignition rate is large. At the end
of this section we discuss other kinds of initial conditions.\\

Please, get the full text of the paper here:\\

www.lehman.edu/faculty/dgaranin/deflagration.pdf

\section{Structure and velocity of the deflagration front}

\label{Sec-Front}

\subsection{Thermodynamics of magnetic deflagration}

Ignition of deflagration leads to a strong increase of the temperature and
relaxation rate that results in equilibration of energy between spin and
phonon subsystems. Since deflagration is a fast process, one can use energy
conservation, neglecting excited states and the heat loss through the
boundaries during deflagration,
\begin{equation}
\mathcal{E}_{i}+n_{-,i}\Delta E=\mathcal{E}_{f}+n_{-}^{(\mathrm{eq}%
)}(T_{f})\Delta E.  \label{EnergyConserv}
\end{equation}
Here $\mathcal{E}_{i,f}\equiv \mathcal{E}(T_{i,f})$ are the phonon energies
at the initial and final temperatures $T_{i}$ (before the deflagration
front) and $T_{f}$ (behind the front), $n_{-,i}$ is the initial population
of the metastable well, and $n_{-}^{(\mathrm{eq})}(T_{f})$ is the
equilibrium value of $n_{-}$ at $T_{f}$ \ given by Eq.\ (\ref{DetailedBalance}%
). We will call $T_{f}$ the \emph{flame temperature}. Eq.\ (\ref
{EnergyConserv}) is a transcedental equation for the flame temperature $%
T_{f}.$ Since $T_{f}$ $\gg T_{i},$ the initial phonon energy $\mathcal{E}%
_{i} $ can be neglected. $T_{f}$ can be found analytically if $\mathcal{E}$
has the form
\begin{equation}
\mathcal{E}=\frac{Ak_{B}\Theta _{D}}{\alpha +1}\left( \frac{T}{\Theta _{D}}%
\right) ^{\alpha +1}  \label{Uph}
\end{equation}
that follows from Eq.\ (\ref{Cph3d}) if $n_{-}^{(\mathrm{eq})}(T_{f})$ is
negligibly small --- the full-burning case. In this case from $\mathcal{E}%
(T_{f})=n_{-,i}\Delta E$ one obtains
\begin{equation}
T_{f}=\Theta _{D}\left( \frac{\left( \alpha +1\right) n_{-,i}\Delta E}{%
Ak_{B}\Theta _{D}}\right) ^{1/(\alpha +1)}.  \label{TfResalpha}
\end{equation}
The full-burning condition is $k_{B}T_{f}\ll \Delta E.$ Using Eq.\ (\ref
{DeltaEDef}) one can rewrite it in the form of the strong-bias condition
\begin{equation}
h\gg h_{\mathrm{fb}}n_{-,i}^{1/\alpha },  \label{hfbDef}
\end{equation}
where the full-burning field
\begin{equation}
h_{\mathrm{fb}}\equiv \frac{k_{B}\Theta _{D}}{4U_{0}}\left( \frac{\alpha +1}{%
A}\right) ^{1/\alpha }
\end{equation}
is a material parameter. With $\alpha =3,$ $A\simeq 234,$ $U_{0}\approx 65$
K and $\Theta _{D}=38$ K for Mn$_{12}$ one obtains
\begin{equation}
h_{\mathrm{fb}}\simeq 0.038.  \label{hfbMn12}
\end{equation}

We will see that the speed of the deflagration front is mainly determined by
the flame's Arrhenius exponent
\begin{equation}
W_{f}\equiv \frac{U}{k_{B}T_{f}}  \label{WfDef}
\end{equation}
that can be large if the energy bias $\Delta E$ is small, resulting in low $%
T_{f}.$ One can express $W_{f}$ in the form
\begin{equation}
W_{f}=\frac{1}{4}\frac{\left( 1-h\right) ^{2}}{\left( n_{-,i}h\right)
^{1/(\alpha +1)}}\frac{1}{h_{\mathrm{fb}}^{\alpha /(\alpha +1)}}.
\label{Wfviahhfb}
\end{equation}
The maximal value of $W_{f}$ compatible with the full-burning condition (for
$n_{-,i}=1)$ is given by
\begin{equation}
W_{f,\max }=\frac{\left( 1-h_{\mathrm{fb}}\right) ^{2}}{4h_{\mathrm{fb}}}%
\approx 6,  \label{Wfmax}
\end{equation}
where the numerical value corresponds to Mn$_{12}$ and uses Eq.\ (\ref
{hfbMn12}). If $W_{f}$ exceeds this value, one cannot neglect $n_{-}^{(%
\mathrm{eq})}(T_{f})$ in determining the flame temperature.

\subsection{The deflagration front}

The deflagration is described by the solution of Eqs.\ (\ref{TnEqsDimensional}%
). Since in molecular magnets at low temperatures both the heat capacity and
thermal conductivity strongly depend on temperature, it is more convenient
to use the phonon energy $\mathcal{E}$ as the dynamical variable instead of $%
T,$ see Eq.\ (\ref{UnEqsDimensional}). The speed of the deflagration front
can be estimated if one rewrites the heat conduction and relaxation
equations in a reduced form
\begin{eqnarray}
\frac{\partial \mathcal{\tilde{E}}}{\partial \tau } &=&\tilde{\nabla}\cdot
\tilde{\kappa}\tilde{\nabla}\mathcal{\tilde{E}}-\frac{\partial \tilde{n}}{%
\partial \tau }  \notag \\
\frac{\partial \tilde{n}}{\partial \tau } &=&-\tilde{\Gamma}(\mathcal{\tilde{%
E}})\tilde{n},  \label{DeflEqsReduced}
\end{eqnarray}
which assumes full-burning. The reduced variables are defined by
\begin{equation}
\mathcal{\tilde{E}}\equiv \frac{\mathcal{E}}{n_{-,i}\Delta E},\qquad \tau
\equiv t\Gamma _{f},\qquad \mathbf{\tilde{r}}\equiv \frac{\mathbf{r}}{l_{d}},
\label{RedVars}
\end{equation}
where $\kappa _{f}$ and $\Gamma _{f}$ are thermal diffusivity and relaxation
rate at the flame temperature $T_{f}$ and
\begin{equation}
l_{d}=\sqrt{\frac{\kappa _{f}}{\Gamma _{f}}}  \label{ldDef}
\end{equation}
describes the width of the deflagration front. The reduced rate is
\begin{equation}
\tilde{\Gamma}(\mathcal{\tilde{E}})\equiv \frac{\Gamma }{\Gamma _{f}}=\exp %
\left[ W_{f}\left( 1-\frac{1}{\tilde{T}(\mathcal{\tilde{E}})}\right) \right]
,  \label{GammatildeDef}
\end{equation}
where $\tilde{\kappa}\equiv \kappa /\kappa _{f},$ and $\tilde{T}\equiv
T/T_{f}$ .

The moving flat deflagration front is a solution of Eqs.\ (\ref
{DeflEqsReduced}) that depends on the combined time-like argument
\begin{equation}
u\equiv \tau -\tilde{x}/\tilde{v},  \label{uFrontDef}
\end{equation}
where $\tilde{v}$ is the reduced deflagration speed. In terms of $u$ Eqs.\ (%
\ref{DeflEqsReduced}) take the form
\begin{eqnarray}
\frac{d\mathcal{\tilde{E}}}{du} &=&\frac{1}{\tilde{v}^{2}}\frac{d}{du}\tilde{%
\kappa}\frac{d\mathcal{\tilde{E}}}{du}-\frac{d\tilde{n}}{du}  \notag \\
\frac{d\tilde{n}}{du} &=&-\tilde{\Gamma}\tilde{n}.  \label{uFrontEqs}
\end{eqnarray}
They represent a nonlinear eigenvalue problem with respect to $\tilde{v}$.
The real deflagration speed $v$ is given by
\begin{equation}
v=\tilde{v}l_{d}\,\Gamma _{f}=\tilde{v}\sqrt{\kappa _{f}\Gamma _{f}}=\tilde{v%
}\sqrt{\kappa _{f}\Gamma _{0}}e^{-W_{f}/2}.  \label{vDef}
\end{equation}

The advantage of using $\mathcal{\tilde{E}}$ as a dynamical variable is that
the first of Eqs.\ (\ref{uFrontEqs}) can be integrated, leading to
\begin{equation}
\tilde{\kappa}(\mathcal{\tilde{E}})\frac{d\mathcal{\tilde{E}}}{du}=\tilde{v}%
^{2}\left( \mathcal{\tilde{E}+}\tilde{n}-1\right) .  \label{EEqIntegrated}
\end{equation}
Far before and far behind the front $\mathcal{\tilde{E}}=\mathrm{const}$ and
one recovers energy conservation, Eq.\ (\ref{EnergyConserv}). One can combine
Eq.\ (\ref{EEqIntegrated}) with the second of Eqs.\ (\ref{uFrontEqs}) to
eliminate $u.$ This results in
\begin{equation}
\tilde{\kappa}(\mathcal{\tilde{E}})\tilde{\Gamma}(\mathcal{\tilde{E}})\tilde{%
n}\frac{d\mathcal{\tilde{E}}}{d\tilde{n}}=-\tilde{v}^{2}\left( \mathcal{%
\tilde{E}+}\tilde{n}-1\right)  \label{EtilntilEq}
\end{equation}
that should be solved with the boundary conditions $\mathcal{\tilde{E}}=0$
at $\tilde{n}=1$ and $\mathcal{\tilde{E}}=1$ at $\tilde{n}=0.$

\subsubsection{Analytical theory}

Eq.\ (\ref{EEqIntegrated}) allows one to find the spatial variation of
temperature before the front, $u<0$ and $|u|\gg a,$ where $\tilde{n}\cong 1.$
For the power-law dependences of Eqs.\ (\ref{Cph3d}) and (\ref{kappaPowerLaw}%
) one has $\tilde{\kappa}(\mathcal{\tilde{E}})=\mathcal{\tilde{E}}^{-\beta
/(\alpha +1)},$ and the solution of Eq.\ (\ref{EEqIntegrated}) yields the
power-law asymptote before the front, $u<0,$
\begin{equation}
\mathcal{\tilde{E}}\cong \left[ \frac{\tilde{v}^{2}\beta }{\alpha +1}\left(
-u+c\right) \right] ^{-(\alpha +1)/\beta },\qquad \tilde{T}\cong \mathcal{%
\tilde{E}}^{1/(\alpha +1)},  \label{etilAstail}
\end{equation}
where $c$ is the integration constant related to the position of the
deflagration front. In the realistic case of $\beta =13/3$ the exponent in
the expression for $\tilde{T}$ is rather small. That is, the heat propagates
far ahead of the deflagration front due to the divergence of thermal
diffisivity at low temperature. In the case of constant $\kappa $ the
temperature before the front decreases exponentially as one moves away from
the front,
\begin{equation}
\mathcal{\tilde{E}}\cong e^{\tilde{v}^{2}u},\qquad \tilde{T}\cong e^{\tilde{v%
}^{2}u/(\alpha +1)}.  \label{EtilAstailExp}
\end{equation}
Here the integration constant additive to $u$ was set to zero$.$ Now one can
find the variation of $\tilde{n}$ before the front from the second of Eqs.\ (%
\ref{uFrontEqs}) and Eq.\ (\ref{etilAstail}), to confirm that $1-\tilde{n}$
is very small.

Behind the front the $T$ is close to the flame temperature, $\tilde{T}\cong
\mathcal{\tilde{E}}\cong 1,$ so that $\tilde{\Gamma}\cong 1$ and from the
second of Eqs.\ (\ref{uFrontEqs}) one obtains
\begin{equation}
\tilde{n}\cong e^{-u}.  \label{ntildeBehind}
\end{equation}
Now one can find the deviation $\delta \mathcal{\tilde{E}}\equiv \mathcal{%
\tilde{E}-}1$ behind the front from Eq.\ (\ref{EEqIntegrated}). Setting $%
\tilde{\kappa}(\mathcal{\tilde{E}})\Rightarrow 1,$ one obtains $\partial _{u}%
\mathcal{\tilde{E}=}\tilde{v}^{2}\left( \mathcal{\tilde{E}+}e^{-u}-1\right) $
that yields
\begin{equation}
\mathcal{\tilde{E}}\cong 1-\frac{\tilde{v}^{2}}{1+\tilde{v}^{2}}e^{-u}=1-%
\frac{\tilde{v}^{2}}{1+\tilde{v}^{2}}\tilde{n}.  \label{Etilbehind}
\end{equation}
This relation between $\mathcal{\tilde{E}}$ and $\tilde{n}$ also can be
obtained from Eq.\ (\ref{EtilntilEq}) with $\tilde{\kappa}(\mathcal{\tilde{E}}%
)\Rightarrow 1$ and $\tilde{\Gamma}\Rightarrow 1$ behind the front.

The speed of the deflagration front can be calculated analytically in the
high-barrier limit $W_{f}\gg 1$. In this case, burning occurs in the region
where the temperature is already very close to the flame temperature, $%
\mathcal{\tilde{E}}\cong 1\mathcal{.}$ Linearizing the argument of the
exponential in Eq.\ (\ref{GammatildeDef}) on $\delta \mathcal{\tilde{E}\equiv
\tilde{E}}-1,$ one obtains for the relaxation rate
\begin{equation}
\tilde{\Gamma}\cong e^{y},\qquad y\equiv \nu _{f}\delta \mathcal{\tilde{E}}
\label{GammaTildeltaEtil}
\end{equation}
with
\begin{equation}
\qquad \nu _{f}\equiv W_{f}\frac{n_{-,i}\Delta E}{C_{\mathrm{ph,}f}T_{f}}.
\label{nufDef}
\end{equation}
The parameter $\nu _{f}$ is similar to $\nu $ of Eq.\ (\ref{nuDef}), only it
is defined with respect to the temperature $T_{f}.$ Using Eqs.\ (\ref{Cph3d})
and (\ref{TfResalpha}), one can simplify $\nu _{f}$ to
\begin{equation}
\nu _{f}=\frac{W_{f}}{\alpha +1}.  \label{nufSimplified}
\end{equation}
According to Eq.\ (\ref{Wfmax}), the maximal value of $\nu _{f}$ compatible
with the full-burning approximation, is $\nu _{f,\max }\approx 1.5.$
Nevertheless, for simplicity we will consider the case $\nu _{f}\gg 1$
within the full-burning approximation. In this case burning occurs only when
the phonon energy is very close to its final value, i.e., $\delta \mathcal{%
\tilde{E}}\sim 1/\nu _{f}\ll 1.$ Hence in Eq.\ (\ref{EEqIntegrated}) one can
make a replacement $\mathcal{\tilde{E}}\Rightarrow 1$ and $\tilde{\kappa}(%
\mathcal{\tilde{E}})\Rightarrow 1$ in the burning region$.$ Eq.\ (\ref
{EtilntilEq}) then takes the form
\begin{equation}
\frac{dy}{d\tilde{n}}=-\nu _{f}\tilde{v}^{2}e^{-y}.  \label{nyEq}
\end{equation}
It is convenient to consider this equation as an equation for $\tilde{n}(y).$
The solution satisfying the boundary conditions $\tilde{n}=1$ before the
front ($y=-\infty )$ and $\tilde{n}=0$ behind the front ($y=0)$ reads
\begin{equation}
\tilde{n}=1-e^{y}.  \label{nyRes}
\end{equation}
It exists if the reduced front speed is given by $\nu _{f}\tilde{v}^{2}=1$
or
\begin{equation}
\tilde{v}=\frac{1}{\sqrt{\nu _{f}}}\ll 1.  \label{vtilRes}
\end{equation}
Note that this result is insensitive to the temperature dependences of the
heat capacity and thermal conductivity. In real units, one obtains from Eq.\ (%
\ref{vDef})
\begin{equation}
v=\sqrt{\frac{\kappa _{f}\Gamma _{0}}{\nu _{f}}}e^{-W_{f}/2},
\label{vDeflHighBarrierFinal}
\end{equation}
where $\nu _{f}$ is given by Eqs.\ (\ref{nufDef}) or (\ref{nufSimplified}).
Now $\delta \mathcal{\tilde{E}}$ can be found from the full system of
equations
\begin{equation}
\frac{dy}{du}=\nu _{f}\tilde{v}^{2}\tilde{n},\qquad \frac{d\tilde{n}}{du}%
=-e^{y}\tilde{n}.  \label{deltaEnminEqs}
\end{equation}
With account of $\nu _{f}\tilde{v}^{2}=1$ and Eq.\ (\ref{nyRes}) the first of
these equations becomes $\partial _{u}y=1-e^{y}.$ The solution is
\begin{equation}
y\mathcal{=-}\ln \left( 1+e^{-u}\right) .  \label{yuRes}
\end{equation}
Then from Eq.\ (\ref{nyRes}) one obtains
\begin{equation}
\tilde{n}=\frac{1}{1+e^{u}}=\frac{1}{2}\left( 1-\tanh \frac{u}{2}\right) .
\label{nuRes}
\end{equation}

The solution for $\mathcal{\tilde{E}}$ in the whole range of $u$ can be
obtained by merging Eqs.\ (\ref{yuRes}) and (\ref{etilAstail}). \ The result
is
\begin{equation}
\mathcal{\tilde{E}=}\frac{1}{\left[ 1+\frac{\beta }{\alpha +1}\tilde{v}%
^{2}\ln \left( 1+e^{-u}\right) \right] ^{(\alpha +1)/\beta }}.
\label{EtilFinal}
\end{equation}
Its accuracy is assured by smallness of $\tilde{v}^{2}.$ In the case of $%
\kappa =\mathrm{const,}$ merging Eqs.\ (\ref{yuRes}) and (\ref{EtilAstailExp}%
) yields
\begin{equation}
\mathcal{\tilde{E}}=\left( 1+e^{-u}\right) ^{-\tilde{v}^{2}}
\label{EtilFinalExp}
\end{equation}
which is the limit of $\beta \rightarrow 0$ in Eq.\ (\ref{EtilFinal}). Note
that the width of the deflagration front $l_{d}$ defined by Eq.\ (\ref{ldDef}%
) is the width of the region where the magnetization changes, see Eq.\ (\ref
{nuRes}). The width of the region where the temperature changes is $l_{d}/%
\tilde{v}=\sqrt{\kappa _{f}\nu _{f}/\Gamma _{f}},$ according to Eqs.\ (\ref
{uFrontDef}) and (\ref{EtilFinalExp}). For thermal diffusivity diverging at $%
T=0$ for $\beta >0,$ the region of the temperature variation becomes very
broad and its width cannot be defined.

\begin{figure}[t]
\unitlength1cm
\begin{picture}(11,5.5)
\centerline{\psfig{file=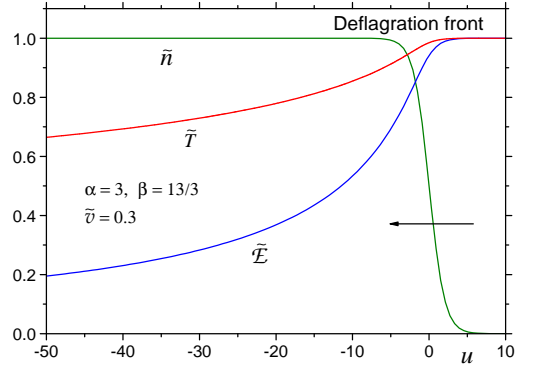,angle=-90,width=8cm}}
\end{picture}
\caption{Analytical results for the deflagration front in molecular magnets
in the high-barrier case, $\tilde{v}=0.3$ (i.e., $\nu_f=1/\tilde{v}^2\simeq 11$). }
\label{Fig-Front-analytical}
\end{figure}
\begin{figure}[t]
\unitlength1cm
\begin{picture}(11,5.5)
\centerline{\psfig{file=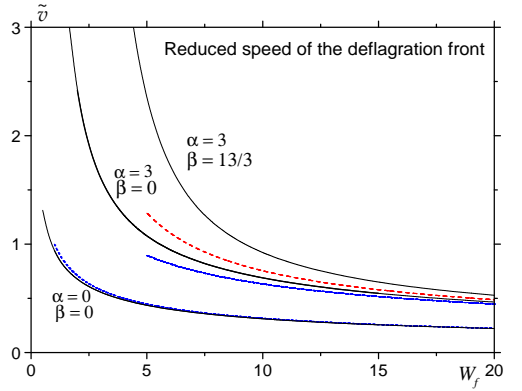,angle=-90,width=8cm}}
\end{picture}
\caption{Numerical results for the reduced deflagration speed $\tilde{v}$ vs
the flame's Arrhenius exponent $W_{f}$ for different temperature dependences
$C_{\mathrm{ph}}\varpropto T^{\protect\alpha }$ and $\protect\kappa %
\varpropto T^{-\protect\beta }.$ Dashed lines are the $W_{f}$ $\gg 1$
asymptotes $\tilde{v}\cong \protect\sqrt{\left( \protect\alpha +1\right)
/W_{f}}.$}
\label{Fig-vtil-vs-Wf}
\end{figure}

\subsubsection{Numerical results for the deflagration front}

If $\nu _{f}$ of Eq.\ (\ref{nufDef}) is not large, the problem of the
deflagration front cannot be solved analytically. Numerical solution uses
Eq.\ (\ref{EEqIntegrated}) and the second of Eqs.\ (\ref{uFrontEqs}). One
starts in the region behind the front, with proper boundary conditions and
arbitrary $\tilde{v},$ and solves equations numerically into the region
sufficiently far ahead of the front$.$ If the value of $\tilde{v}$ is
correct, and only in this case, the solution ahead of the front will be $%
\tilde{n}=1$ and $\mathcal{\tilde{E}}$ having the asymptotic form of Eq.\ (%
\ref{etilAstail}). One finds $\tilde{v}$ numerically from one of these
conditions, with consistent results. The numerically calculated dependences
of $\tilde{v}$ on $W_{f}$ are shown in Fig.\ \ref{Fig-vtil-vs-Wf}.
Surprisingly, the high-barrier analytical theory works very well in the
whole range of $W_{f}.$

To the contrary, the validity of the analytical theory in the realistic case
of $\alpha =3$ and $\beta =13/3$ requires rather large $W_{f}.$ One of the
reasons for this is that the large-$W_{f}$ approximation in fact requires
large $\nu _{f}$ in Eq.\ (\ref{nufSimplified}). Nonzero exponents $\alpha $
and $\beta $ have the following effect on the deflagration speed. For $%
\alpha >1$ the heat-conduction equation is written in terms of $\mathcal{%
\tilde{E}.}$ The decrease of $\mathcal{\tilde{E}}$ ahead of the front leads
to the decrease of $\tilde{T}(\mathcal{\tilde{E}})=\mathcal{\tilde{E}}%
^{1/(\alpha +1)}$ that enters the relaxation rate $\tilde{\Gamma}.$ For $%
\alpha >0$ this temperature decrease is less essential than in the case of $%
\alpha =0.$ Thus the temperature before the front is higher and the front
moves faster because of faster relaxation. The role of $\beta >0$ is
similar. The heat diffision in the region before the front is faster, the
temperature before the front is higher, and the deflagration speed increases.

\section{Full numerical solution of the deflagration problem}

\label{Sec-Numerical}

Please, get the full text of the paper here:\\

www.lehman.edu/faculty/dgaranin/deflagration.pdf

\section{Discussion}

\label{Sec-Discussion}

So far experimental work on magnetic deflagration has been limited to
measurements of the flame speed. The proposed theory provides the framework
for more detailed experimental studies suggested in this section.

\subsection{Deflagration threshold}

According to the theory the magnetic deflagration in a crystal of magnetic
molecules can be ignited by either increasing magnetic field or temperature.
The simplest situation is when the magnetic field and temperature of the
sample boundary $T_{0}$ are independent of coordinates. In this case the
crystal loses stability against formation and propagation of the flame
(magnetic avalanche) when the rate of the spin flip for an individual
molecule, $\Gamma (H,T_{0})$, exceeds
\begin{equation}
\Gamma _{c}=\frac{8k(T_{0})k_{B}T_{0}^{2}}{U(H)\Delta E(H)n_{-,i}l^{2}}\,.
\label{threshold}
\end{equation}
Here $k(T_{0})$ is coefficient of thermal conductivity at $T=T_{0}$, $U(H)$
and $\Delta E(H)$ are field-dependent energy barrier and energy difference
respectively between spin-up and spin-down states, $l$ is some
characteristic length, and $n_{-,i}$ is the initial fraction of molecules
available for burning. It can be expressed via (negative) initial
magnetization $M_{i}$ and the saturation magnetization $M_{0}$ as $%
n_{-,i}=(M_{0}-M_{i})/(2M_{0})$. As to the parameter $l$, it is uniquely
determined by geometry and is of order of the smallest dimension of the
crystal. Eq.\ (\ref{threshold}) provides the dependence of the critical
magnetic field on the temperature of the sample, or, inversely, the
dependence of the critical temperature of the sample on the magnetic field,
see Fig.\ \ref{Fig-QDT}. The deeps in $T_{0}(H)$ at regularly spaced fields
are due to the maxima of $\Gamma (H,T_{0})$ at tunneling resonances. Note
that Eq.\ (\ref{threshold}) contains explicit dependence of the deflagration
threshold on the initial magnetization that should be easy to test in
experiment.

Similar relations have been obtained by us in the presence of field and
temperature bias, see Sec.\ \ref{Sec-threshold}. Among other things we have
demonstrated that the bias suppresses deflagration. The most important
outcome of this studies is elucidation of the nature of magnetic avalanche.
Contrary to the initial beliefs, the avalanche does not develop from a small
nucleus of the magnetization reversal inside the crystal. It begins as an
instability of a smooth temperature profile when the spin-flip rate of
individual molecules (the burning rate) exceeds the rate at which the heat
flows out of the burning region. The effect is exponentially sensitive to
the magnetic field and temperature of the sample. Even a slight fluctuation
of $H$ or $T_{0}$ may take the system deep inside the instability region ($%
\Gamma \gg \Gamma _{c}$), thus, explaining the abrupt and sometimes
unpredictable nature of the avalanche. It will be interesting to see if
experiments confirms our predictions for the deflagration threshold at
various initial conditions.

\subsection{Ignition time}

In this paper we have addressed situations when the crystal is
instantaneously brought inside the instability region. In Sec.\ \ref
{Sec-Ignition-rate} we have demonstrated that the ignition of the
deflagration occurs after a finite time elapses from the moment when the
instability threshold is crossed. When crossing the threshold on field or
temperature the ignition rate changes from zero below the threshold to a
some finite value above the threshold, see Figs.\ \ref
{Fig-IgnitionExplosive} -- \ref{Fig-Ignition-EdgeRise}. The deeper one
penetrates into the instability region the smaller is the ignition time.
Most of the experiments on magnetic avalanches were done in a field-sweep
mode, when the magnetic field changes at a constant rate, $H=rt$, from a
large negative value to a large positive value. In this case, the field $%
H_{c}^{\prime }$ at which the avalanche occurs should be approximately
determined by the equation $H_{c}^{\prime }=H_{c}+r\tau _{ig}(H_{c}^{\prime
})$, where $H_{c}$ is the critical field at a temperature $T_{0}$ plotted in
Fig.\ \ref{Fig-QDT}. Since the ignition time rapidly falls as $H$ grows above
$H_{c}$, it is clear that for sufficiently small sweep rates $r$, the field $%
H_{c}^{\prime }$ must be very close to $H_{c}$. This condition is always
fulfilled in a field-sweep experiment unless a pulse field technique is used
with a very large $r$. Thus, a typical field-sweep experiment is capable of
testing the $H_{c}(T_{0})$ dependence plotted in Fig.\ \ref{Fig-QDT}, but not
probing the ignition time. To measure the ignition time one should apply
different techniques. The trick is to cross the stability threshold by a
finite step on field or temperature during the time interval that is small
compared to the ignition time. This can be achieved by using a small coil
with a short time constant in addition to the large coil needed to bring the
system close to the threshold. Alternatively, one can use fast heaters to
cross the deflagration threshold on the temperature of the sample or on the
temperature of one end of the sample along the lines of Section \ref
{Sec-Ignition-rate}c.

Results obtained in one dimension (Sections \ref{Sec-Ignition-rate} and \ref
{Sec-Numerical}) show that under symmetric conditions the avalanche ignites
in the middle of the sample or simultaneously at two symmetric regions away
from the center. Symmetry arguments suggest that this should also be the
case for any symmetric sample. Meantime experiments done under uniform field
and temperature conditions often report the ignition of the deflagration
persistently at one end of the sample. Explanation of this observation
should be sought in the inevitable asymmetry of the heat flow inside and out
of the real crystal. Such an asymmetry occurs due to the asymmetry of the
crystal shape, internal inhomogeneity, or as a result of the asymmetric
thermal insulation from the environment. It should lead to the asymmetric
temperature profile of a quasi-equilibrium state below the deflagration
threshold and, thus, asymmetric ignition of the deflagration. Our studies
elucidate the crucial role of the boundary conditions. This should be
addressed in future experiments by studying, e.g., deflagration in thermally
insulated crystals alongside with crystals that freely exchange heat with
the environment, and under asymmetric boundary conditions.

\subsection{Velocity and width of the deflagration front}

Measured field and temperature dependences of the velocity of the magnetic
avalanche are in a reasonably good agreement with the concept of
deflagration. \cite{suzetal05prl,heretal05prl} This agreement, however, has
only been established with an accuracy to the exponent, the prefactor was
estimated by order of magnitude. In Section \ref{Sec-Front} we provided a
more detailed study of the developed deflagration. Our result for the speed
of the deflagration front reads
\begin{equation}
v(H)=\sqrt{\frac{4k_{B}T_{f}\kappa (T_{f})\Gamma (H,T_{f})}{U(H)}}\,,
\end{equation}
where $\kappa $ is thermal diffusivity,
\begin{equation}
T_{f}=\frac{\Theta _{D}}{\pi }\left[ \frac{5}{3}\frac{n_{-,i}\Delta E(H)}{%
k_{B}\Theta _{D}}\right] ^{1/4}
\end{equation}
is the flame temperature (the temperature behind the front), and $\Theta _{D}
$ is the Debye temperature. Numerical exercise with numbers for Mn$_{12}$
and fields used in experiment immediately shows that the above formulas give
correct estimate of $v$ and $T_{f}$ and their correct field dependence.
Future experiments should show whether these formulas provide quantitative
description of the developed magnetic deflagration.

An interesting observation that follows from our theory is that in the
developed magnetic deflagration the width of the region where magnetization
reverses is different from the width of the region inside which the
temperature decays from $T_{f}$ to $T_{0}$. In fact, the latter region is
very broad and even difficult to define, see Fig.\ \ref{Fig-Front-analytical}%
. This is a result of the divergence of thermal diffusivity at $T\rightarrow
0$, which makes low-temperature magnetic deflagration different from
chemical deflagration. The latter has a well-defined width, $l_{d}\sim \sqrt{%
\kappa (T_{f})/\Gamma (T_{f})}$. In the magnetic case, however, this formula
applies only to the width of the region where the magnetization reverses,
but not to the region where the temperature changes. This may explain
reported difficulties in local measurements of the temperature during the
deflagration process.

\section{Acknowledgements}

We thank members of experimental groups of Myriam Sarachik and Javier Tejada
for many fruitful discussions and for providing us with experimental data.
This work has been supported by the NSF Grant No. EIA-0310517.

\bibliographystyle{plain}
\bibliography{gar-own,chu-own,gar-tunneling,gar-books}

\begin{thebibliography}{10}

\bibitem{heretal05prl}
{A. Hern\'andez-Minguez, J. M. Hern\'andez, F. Macia, A. Garcia-Santiago, J.
  Tejada, P. V. Santos}.
\newblock Quantum magnetic deflagration in {M}n$_{12}$ acetate.
\newblock {\em Phys. Rev. Lett.}, 95:217205--(4), 2005.

\bibitem{gometal98prb}
{A. M. Gomes, M. A. Novak, R. Sessoli, A. Caneschi, and D. Gatteschi}.
\newblock Specific heat and magnetic relaxation of the quantum nanomagnet
  {M}n$_{12}$-{A}c.
\newblock {\em Phys. Rev. B}, 57:5021, 1998.

\bibitem{gomnovnunrap01jmmm}
{A. M. Gomes, M. A. Novak, W. C. Nunes, and R. E. Rapp}.
\newblock {\em J. Magn. Magn. Mater.}, 226-230:2015, 2001.

\bibitem{kit63}
{C. Kittel}.
\newblock {\em Quantum {T}heory of {S}olids}.
\newblock Wiley and Sons, New York -- London, 1963.

\bibitem{paupark95kluwer}
{C. Paulsen and J.-G. Park}.
\newblock In L.~Gunther and B.~Barbara, editors, {\em Quantum Tunneling of
  Magnetization -- {QTM}'94}. Kluwer, Dordrecht, 1995.

\bibitem{garchu97prb}
{D. A. Garanin and E. M. Chudnovsky}.
\newblock Thermally activated resonant magnetization tunneling in molecular
  magnets: Mn$_{12}${A}c and others.
\newblock {\em Phys. Rev. B}, 56:11102--11118, 1997.

\bibitem{garlut92ap}
{D. A. Garanin and V. S. Lutovinov}.
\newblock Absorption of sound and kinetic coefficients of elastic bodies.
\newblock {\em Ann. Phys. (N.Y.)}, 218:293--324, 1992.

\bibitem{baretal99prb}
{E. del Barco, J. M. Hern\'andez, M. Sales, J. Tejada, H. Rakoto, J. M. Broto,
  and E. M. Chudnovsky}.
\newblock Spin-phonon avalanches in in {M}n-12 {A}cetate.
\newblock {\em Phys. Rev. B}, 60:11898--11901, 1999.

\bibitem{chutej98book}
{E. M. Chudnovsky and J. Tejada}.
\newblock {\em Macroscopic quantum tunneling of the magnetic moment}.
\newblock Cambridge University Press, Cambridge, 1998.

\bibitem{chutej06book}
{E. M. Chudnovsky and J. Tejada}.
\newblock {\em Lectures on {M}agnetism}.
\newblock Rinton Press, Princeton, 2006.

\bibitem{fometal97prl}
{F. Fominaya, J. Villain, P. Gaudit, J. Chaussy, and A. Caneschi}.
\newblock Heat capacity anomalies induced by magnetization quantum tunneling in
  {M}n$_{12}${O}$_{12}$-acetate single crystal.
\newblock {\em Phys. Rev. Lett.}, 79:1126--1129, 1997.

\bibitem{fometal99prb}
{F. Fominaya, J. Villain, T. Fournier, P. Gandit, J. Chaussy, A. Fort, and A.
  Caneschi}.
\newblock Magnetic-field-dependent thermodynamics of {M}n$_{12}$ acetate single
  crystals at low temperatures.
\newblock {\em Phys. Rev. B}, 59:519, 1999.

\bibitem{gla96book}
{I. Glassman}.
\newblock {\em Combustion}.
\newblock Academic Press, 1996.

\bibitem{feralo05prb}
{J. F. Fern\'andez and J. J. Alonso}.
\newblock Time relaxation of interacting single-molecule magnets.
\newblock {\em Phys. Rev. B}, 72:094431--(8), 2005.

\bibitem{heretal96epl}
{J. M. Hern\'andez, X. X. Zhang, F. Luis, J. Bartolom\'e, J. Tejada, and R.
  Ziolo}.
\newblock Field tuning of thermally activated magnetic quantum tunneling in
  {Mn$_{12}$Ac} molecules.
\newblock {\em Europhys. Lett.}, 35:301--306, 1996.

\bibitem{frisartejzio96prl}
{J. R. Friedman, M. P. Sarachik, J. Tejada, and R. Ziolo}.
\newblock Macroscopic measurement of resonant magnetisation tunneling in
  high-spin molecules.
\newblock {\em Phys. Rev. Lett.}, 76:3830--3833, 1996.

\bibitem{sesgatcannov93nat}
{R. Sessoli, D. Gatteschi, A. Caneschi, and M. A. Novak}.
\newblock Magnetic bistability in a metal-ion cluster.
\newblock {\em Nature (London)}, 365:141, 1993.

\bibitem{suzetal05prl}
{Y. Suzuki, M. P. Sarachik, E. M. Chudnovsky, S. McHugh, R. Gonzalez-Rubio, N.
  Avraham, Y. Myasoedov, E. Zeldov, H. Shtrikman, N. E. Chakov, and G.
  Christou}.
\newblock Propagation of avalanches in {M}n$_{12}$-{A}cetate: {M}agnetic
  deflagration.
\newblock {\em Phys. Rev. Lett.}, 95:147201--(4), 2005.

\end{thebibliography}

\end{document}